\documentclass[epj,final]{svjour}

\usepackage{graphics,bbm}
\begin{document} 
  \title{Similarities between the $t-J$ and Hubbard models in weakly 
correlated regimes}

  \author{Marie-Bernadette LEPETIT\inst{1} \and Marie-Liesse
  DOUBLET\inst{2} \and Philippe MAUREL\inst{1}}

    \institute{Laboratoire de Physique Quantique, IRSAMC, Université
    Paul Sabatier, 118 Route de Narbonne, 31062 Toulouse, France \and
    Laboratoire de Structure et Dynamique des Systèmes Moléculaires et
    Solides USTL II, Bât. 15 CC-014, 34095 Montpellier C\'edex~5,
    France}

  \date{\today} 


\abstract{
We present a comparative study of the Hubbard and $t-J$ models far
away from half-filling. We show that, at such fillings the $t-J$
Hamiltonian can be seen as an effective model of the repulsive Hubbard
Hamiltonian over the whole range of correlation strength. Indeed, the
$|t/U| \in \left[0,+\infty \right[$ range of the Hubbard model can be
mapped onto the finite range $|J/t| \in \left[1, 0 \right]$ of the
$t-J$ model, provided that the effective exchange parameter $J$ is
defined variationally as the local singlet-triplet excitation energy.
In this picture the uncorrelated limit $U=0$ is associated with the
super-symmetric point $J=-2|t|$ and the infinitely correlated
$U=+\infty$ limit with the usual $J=0$ limit. A numerical comparison
between the two models is presented using different macroscopic and
microscopic properties such as energies, charge gaps and bond orders
on a quarter-filled infinite chain. The usage of the $t-J$ Hamiltonian
in low-filled systems can therefore be a good alternative to the
Hubbard model in large time-consuming calculations.
\PACS{{71.10-w}{Theories and models of many-electron systems 
} \and {71.10.Fd}{Lattice fermion models}}
}

\maketitle


\section{Introduction} 

In the last decade a renewed interest has been observed, in the study
of simple models such as the Heisenberg model~\cite{heis},
Hubbard~\cite{hub} or extended Hubbard model, the $t-J$
model~\cite{tj}, etc. This attraction is related to the synthesis by
solid state chemists in the last two decades, of a large number of
strongly correlated systems presenting very attractive low energy
properties, often directly linked to the correlation effects. It is
enough to say that these simple models are considered pertinent for
the description of systems such as organic conductors~\cite{bech} or
high $T_c$ super-conductors~\cite{tj} to understand the importance of
their study.

Among these models the Hubbard (or extended Hubbard) model is often
considered as a reference since it has been built to reproduce 
the physics over the whole range of correlation strength,
from delocalized to strongly correlated systems. The other models, for
instance the spin models, are then considered as effective
Hamiltonians of the Hubbard one under certain conditions. In
particular the spin models, Heisenberg and $t-J$ are unanimously
recognized as effective description of the valence physics in the
large correlation limit ($U/t \longrightarrow \infty$ where $U$ is the
on site Coulomb repulsion and $t$ the nearest neighbor hopping
integral). Indeed, in a half-filled, one-band Hubbard model
$$ H_{Hub} = t \sum_{<i,j>} \sum_\sigma{\left(
a^\dagger_{i,\sigma}a_{j,\sigma} +
a^\dagger_{j,\sigma}a_{i,\sigma}\right) } + U \sum_i
n_{i,\uparrow}n_{i,\downarrow}$$ (where $a^\dagger_{i,\sigma}$,
$a_{i,\sigma}$ and $n_{i,\sigma}$ are the usual creation, annihilation
and number operators of an electron of spin $\sigma$ on site $i$, and
$<i,j>$ symbolizes sites linked by the delocalization process), the
probability of a site double occupancy rapidly decreases as $4t^2/U^2$,
in the large $U/t$ regime.  The valence configurations involving sites
double-occupancies are therefore negligible and can be excluded from
an explicit representation, provided that their effects on the other
configurations (involving only singly-occupied sites  or vacant
sites) is reproduced.  As it is well known, this is exactly what is
achieved by the Heisenberg Hamiltonian
\begin{eqnarray*}
H_{Heis}& =& -J\sum_{<i,j>}P\left({\bf S_i .  S_j} - 1/4 \; n_in_j
\right)P \\
&=&  -J  \sum_{<i,j>} P
\left( {a^\dagger_{i,\uparrow}a^\dagger_{j,\downarrow} - 
a^\dagger_{i,\downarrow}a^\dagger_{j,\uparrow} \over \sqrt{2}} \right)
\left( {a^{\rule{0mm}{1ex}}_{i,\uparrow}a_{j,\downarrow} - 
a_{i,\downarrow}a_{j,\uparrow} \over \sqrt{2}} \right) P \\
&=& -J\, \sum_{<i,j>} PS\!g^{\dagger}(i,j) S\!g(i,j)P
\end{eqnarray*}
where $S\!g^\dagger(i,j)$ (resp. $S\!g(i,j)$) is creating
(annihilating) a singlet on the $i,j$ bond and $P=\prod_{i}\left(
\mathbbmss{1} - n_{i\uparrow}n_{i\downarrow} \right)$ is the projector
over all configurations excluding the double occupancy of a site.  \\
Doping the system in holes (resp. in electrons) adds a new
delocalization possibility for the particles which can thus be
described by the so-called $t-J$ model
\begin{eqnarray*} \hspace*{-5eM}
 H_{t-J} &=& t \sum_{<i,j>} \sum_\sigma
Pa^\dagger_{i,\sigma}a_{j,\sigma}P -J \sum_{<i,j>}P\left({\bf S_i . S_j}
- 1/4\;n_in_j \right)P \\
&=& t \sum_{<i,j>} \sum_\sigma
Pa^\dagger_{i,\sigma}a_{j,\sigma}P \\&& -J \sum_{<i,j>}
P\left( {a^\dagger_{i,\uparrow}a^\dagger_{j,\downarrow} - 
a^\dagger_{i,\downarrow}a^\dagger_{j,\uparrow} \over \sqrt{2}} \right)
\left( {a^{\rule{0mm}{1ex}}_{i,\uparrow}a_{j,\downarrow} - 
a_{i,\downarrow}a_{j,\uparrow} \over \sqrt{2}} \right)P
\end{eqnarray*}
This simple derivation is very well known and the $t-J$ model which is
thus considered as the strongly correlated effective Hamiltonian
of the Hubbard model away from half-filling.

In this paper we would like to come back on this assumption and show
that the $t-J$ model can be seen as an effective model for the Hubbard
Hamiltonian over the whole range of correlation strength, provided
that one is far enough from half-filling.

The next section will be devoted to the physical justification and
analytical development of the above assumption using the effective
Hamiltonian theory~\cite{heff}. Section three will assert the validity
of the $t-J$ versus Hubbard equivalence for low-filling systems
through numerical comparisons (energies, gaps, bond-orders) of a
quarter-filled dimerized chain in the Hubbard and $t-J$ models, for the
whole range of correlation and dimerization strengths.  Eventually the
last section will discuss the limitations of an effective
Hamiltonian reduced to two-bodies interactions.

\section{A variational effective exchange integral}

We would like to start by drawing the attention of the reader on the
reasons supporting the validity of the Heisenberg Hamiltonian as an
effective model of the Hubbard Hamiltonian in the strongly correlated
limit. The crucial point is that this is not not the value of the
correlation strength which is important, but rather the fact that the
probability of sites double-occupancies is very small in the ground
state and low lying excited states wave functions. The relevance of
this distinction appears mainly away from half-filling.  Indeed, while
in half-filled systems the low probability of the double occupancy is
a direct consequence of the large correlation, far from half-filling
this is no longer the case. Actually, for low-filled systems (and
equivalently by hole particle symmetry for nearly-filled systems) the
configurations associated with site(s) double-occupancies are
negligible independently of the correlation strength. One can indeed
verify that even in the non interacting regime ($U=0$), where the
double occupancies on a site are the most probable over the whole
range of correlation strength, the wave-function of a $\eta$-filled
system yields a statistical weight for double occupancies on a site of
$\eta^2$. Although it does correspond to a quite-large contribution at
half-filling ($\eta^2=1/4$), it becomes rapidly very small when the
filling is reduced, with a value of only $\eta^2=1/16=0.0625$ for the
quarter-filling case. Since the electron-electron correlation can only
reduce this number, it seems justified to exclude the explicit
reference to double-occupancies in the wave-function of low-filled (or
nearly-filled, using the hole/particle symmetry) systems. However, as
in the Heisenberg limit, simply projecting out the double occupancies
on a site would not give the correct physics and the effects of these
double occupancies on the other states should be effectively
reproduced. As we have seen, this is exactly the role of the effective
exchange integral in the Heisenberg model as well as the $t-J$ model,
that is to lower the local singlet states compared to the local
triplet ones, setting the energy of the neutral representation of a
local singlet $\left(|a\bar{b}\rangle -
|\bar{a}b\rangle\right)/\sqrt{2} \quad {\rm to} \quad -J$ compared to
the local triplet $\left(|a\bar{b}\rangle +
|\bar{a}b\rangle\right)/\sqrt{2}$ that lies at $0$. In the strongly
correlated limit the singlet-triplet excitation energy is evaluated
perturbatively. However, since we would like to derive an effective
model valid over the whole range of correlation strength, perturbative
evaluations of the effective exchange have to be excluded. We will
therefore use locally the principle of the effective Hamiltonian
theory~\cite{heff} (reproduction by the effective Hamiltonian of the
exact eigen-energies and projection of the exact eigen-states over the
model space~: $H_{e\!f\!f} P\Psi_{e\!x\!a\!c\!t} = E_{e\!x\!a\!c\!t}
P\Psi_{e\!x\!a\!c\!t}$) in order to derive a variational evaluation of
the main effects of double-occupancies on a site (from now on denoted
as DOC) on the low energy states (for a more detailled description one can refer to ref.~\cite{ml1}).

Summarizing the previous discussion we can conclude that far away from
half-filling, the configurations including site(s) double-occupancies
can be excluded from an explicit treatment, provided that their main
effect, the lowering of the local singlets compared to the local
triplets, are reproduced by an effective exchange integral. We
therefore see that the Hubbard Hamiltonian can be modeled by the $t-J$
Hamiltonian in the low-filling regime, providing a non perturbative
effective exchange integral, $J(t,U)$. The simplest way to derive
$J(t,U)$ is to impose that it reproduces the variational local
singlet-triplet energy difference on a bond (two sites, two
electrons). The Hubbard model yields $E(S\!g) = \left( U-\sqrt{U^2 + 16t^2}
\right) / 2$, $E(T\!p)= 0 $, it comes
\begin{eqnarray} \label{eq:j}
J(t,U) &=& {U-\sqrt{U^2 + 16t^2} \over 2} 
\end{eqnarray}
Let us note that we retrieve for large values of $t/U$ the
perturbative evaluation of $J$~: $-4t^2/U$ as the first term of the
Taylor expansion in $t/U$.
For the non-correlated limit the expansion should be done in terms of
$U/t$ and it yields
$$ J = {U - \sqrt{U^2+16t^2} \over 2} \simeq -2|t| $$ that is the
value of $J$ for the super-symmetric point.  The super-symmetric point
of the $t-J$ model (the only one to be exactly soluble~\cite{ssexact})
therefore appears as an effective description ---~on the model space
projecting out all site(s) double-occupancies~--- for the
tight-binding Hamiltonian. In this perspective the results of Yokoyama
{\em et al}~\cite{fll} about the super-symmetric point properties
showing that it realizes a Fermi-Liquid state or free Luttinger
liquid and behaves in all points as a free-electrons gas seem quite
natural.  Meanwhile, the effective Hamiltonian theory states that the
ground-state wave-function of a properly defined effective Hamiltonian
should be the projection over the model space of the exact Hamiltonian
one~\cite{heff}.  This property agrees nicely with the results of
Yokoyama {\it et al}~\cite{yoko} that shows that the G\"utzwiller
state is the ground state of the $t-J$ model at the super-symmetric
point. In other words the projection of the free-electron
wave-function onto the space excluding all double occupations on a
site is the $t-J$ ground state for $J(t,U=0)$.

We would like to point out that the effective exchange $J(t,U)$ valid
to represent the repulsive Hubbard model is boundered and varies from
$J(t,U=0)=-2|t|$ for the non correlated limit to $J(t,U=\infty)=0$ for
the strongly correlated limit.  Let us note that these boundaries
excludes the whole range of parameters for which phase separation
phenomena occur. It should also be noted that the arguments leading to
the effective exchange $J(t,U)$ is not specific to the dimension one
and should be expected to hold in all dimensions, at least for non
frustrated systems.

Apart from the effective exchange effect, it is known from
perturbative expansion at large correlation strength, that the DOC
induce effective 3 bodies interactions of lesser importance.This point 
will be discussed in section four and a way to obtain a
variational evaluation of these terms will be proposed.

\section{The dimerized quarter-filled chain}

In this section we will show numerically that the equivalence between
the Hubbard and $t-J$ models really holds for all values of the
correlation strength and for all possible dimerizations.  For this
purpose we will compare the following low energy properties of a
quarter-filled chain in the two models.
\begin{itemize}
\item The behavior of the total energy per site as a function of $t/U$
and the dimerization amplitude $\delta$. This point can be crucial for
a good representation of phase transitions in systems presenting a
more complex topological graph.
\item The charge gap. This criterion will attest for a good low energy
spectroscopic behavior.
\item The effective bond order, as it testifies for the wave function
behavior.
\end{itemize}

The calculations will be done using the infinite systems Density
Matrix Renormalization Group ($DMRG$) method~\cite{white}. We choose
the number of states kept in the renormalized blocks in such a way
that for each model the $10$-sites system is treated exactly. The
Hubbard model will therefore be described with $256$ states and the
$t-J$ one with only $100$ states. These numbers of states kept per
renormalized block correspond to a sum of the discarded states weights
always smaller than $1\times 10^{-7}$, for the Hubbard model, and
smaller than $1\times 10^{-5}$, for the $t-J$ model. These results are
for small dimerization values and small correlation values, that is
for the most difficult cases in terms of the DMRG algorithm
convergence and precision. These small values of the discarded states
weight insures the good convergence of our results, even-though we did
use  the infinite size DMRG algorithm.

The dimerization amplitude is defined as 
$$\delta = \left(t_l - t_w \right)/ t_l$$ where $t_l$ is the large
hopping integral and $t_w$ is the weak one.

\subsection{Energies per site}

\begin{center}
\begin{figure}[h]
\centerline{\resizebox{5cm}{5cm}{\includegraphics{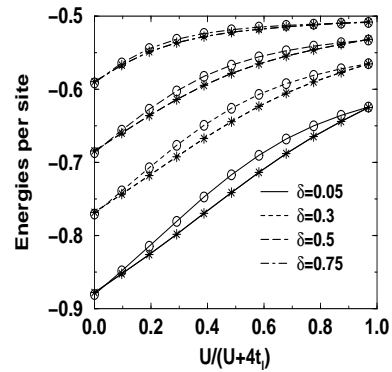}}\vspace{1cm}}
\caption{Energies per site as a function of $U/(U+4|t_l|)$ and
$\delta$ for the Hubbard (stars) and $t-J$ (circle) models, in $t_l$
units.}
\label{fig:esite_Hub-tJ}
\end{figure}
\end{center}

The energies per site have been extrapolated with respect to the
number of sites in the chain and are presented  on
figure~\ref{fig:esite_Hub-tJ} as a function of the correlation
strength $U/(U+4 \mid t_l \mid)$. Different values of the dimerization
factor $\delta$ have been investigated starting from the weakly
dimerized chain $\delta=0.05$ up to the strongly dimerized one
$\delta=0.75$.

We see on figure~\ref{fig:esite_Hub-tJ} that the agreement between the
two models is quite good. Both the variations as a function of the
correlation strength and the dimerization amplitude are well
reproduced. The maximal relative error between the Hubbard and $t-J$
total energies per site is only $0.03$.  As expected, the large $U/\mid
t_l \mid$ asymptotic limit is well reproduced by the $t-J$ model for
all values of the dimerization. A little less obvious is the very good
numerical accuracy found for the non-interacting limit. The relative
error found between our computed $t-J$ values and the {\em exact} free
electron solution is bounded by $0.004$ and can probably be imputed
for its greatest part to the DMRG procedure (which is known to have
its worse convergence properties for the non correlated limit). Actually,
the largest errors occur in the region of intermediate
correlation strength ($U/t\simeq4$) and for small dimerization.  As
the dimerization increases and for a given $U/t$ ratio, the $t-J$ model
fits better and better the Hubbard one.  This point can be easily
understood if one remembers that the three bodies terms, neglected in
the $t-J$ Hamiltonian, should decrease as a function of the
dimerization amplitude. 

\subsection{Charge Gaps}

\begin{figure}[h]
\begin{center}\vspace{2eM}
\resizebox{5cm}{5cm}{\includegraphics{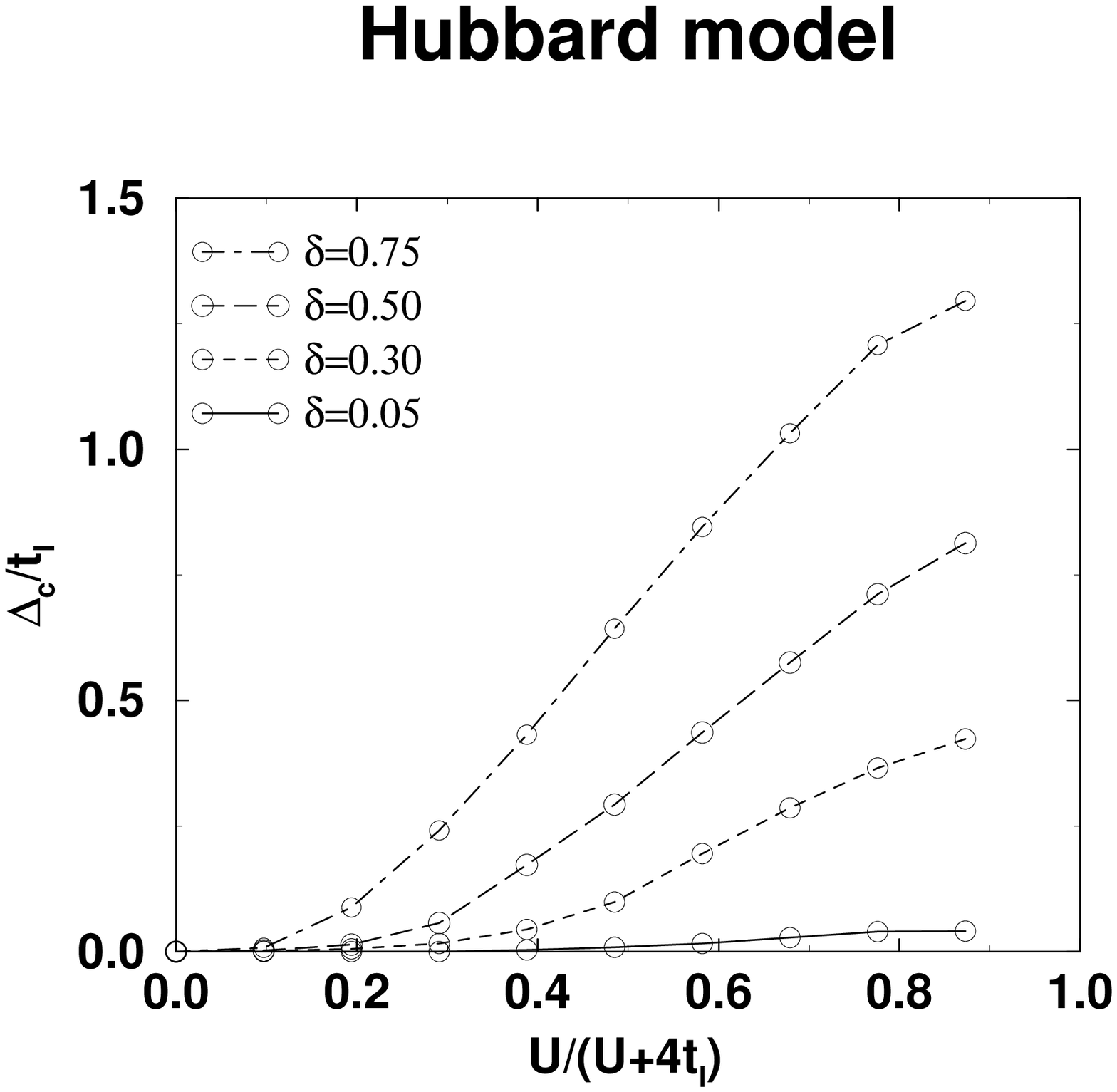}} \\ \vspace{2eM}
\resizebox{5cm}{5cm}{\includegraphics{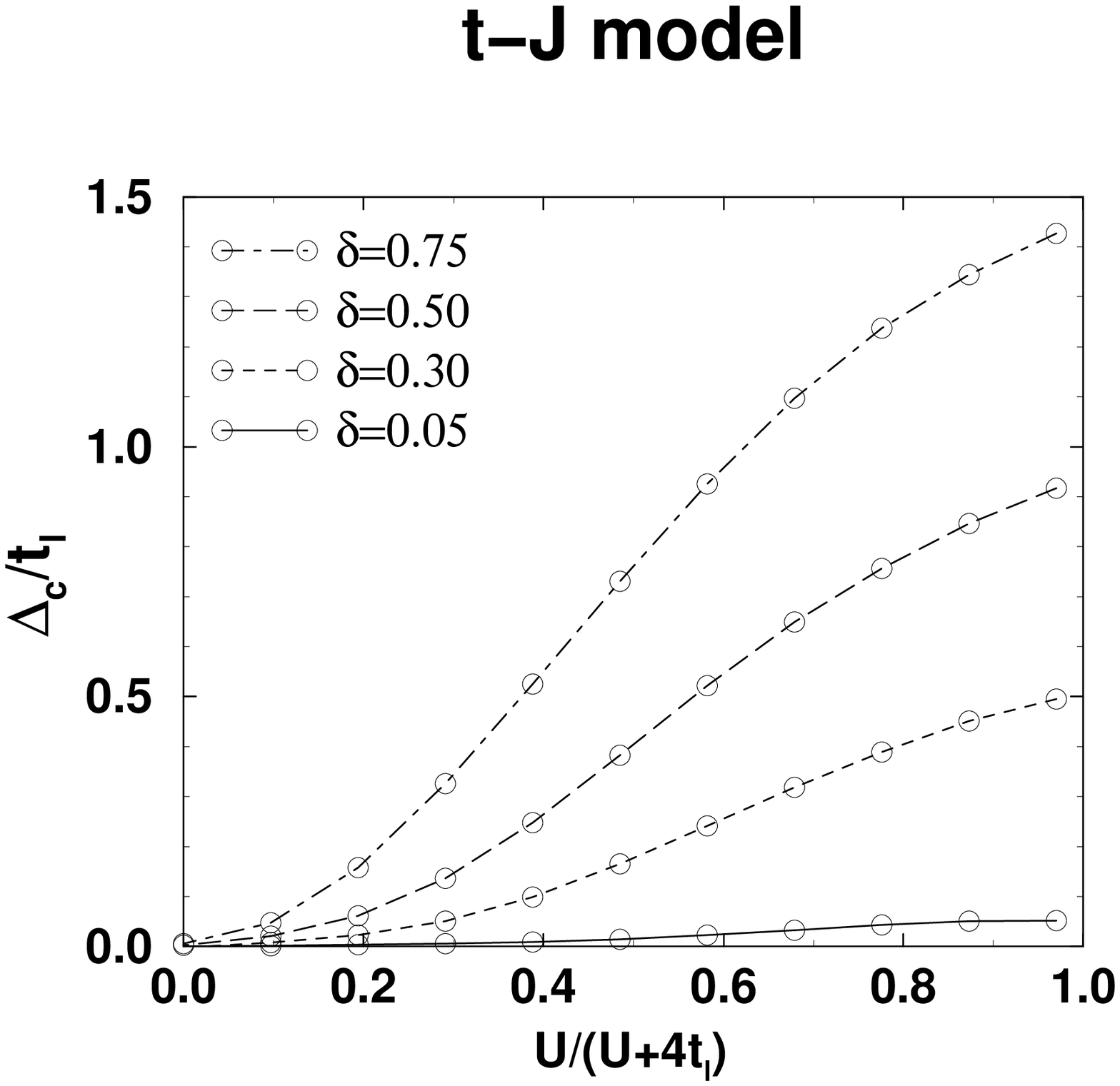}}
\caption{Charge gaps $\Delta_\rho$ in $t_l$ units calculated within
the Hubbard and $t-J$ models. Each curve corresponds to a different
value of $\delta$, going from the weakly dimerized limit $\delta=0.05$
to the strongly dimerized one $\delta=0.75$}
\label{fig:gap}
\end{center}
\end{figure}

We computed the charge gaps $\Delta_\rho$, as the extrapolated
difference toward the infinite system limit, between the ionization
potential (IP) and the electron affinity (EA). 
If $E(N_{site},N_e)$ is the energy
of a finite $N_{site}$ sites system with $N_e$ electrons, the
charge gap is written as the following :
\begin{eqnarray*}
\Delta_\rho &=& \lim_{N_{site}\longrightarrow \infty} \left\{
E(N_{site};N_{site}/2+1)\right. + \\ && \left. E(N_{site};N_{site}/2-1) -
2E(N_{site};N_{site}/2) \right\}
\end{eqnarray*}
The Hubbard and $t-J$ gaps  are presented on figure~\ref{fig:gap} 
 as a function of the correlation strength 
$U/(U+4 \mid t_l \mid)$, and for 4 different values of $\delta$.

One sees immediately that the dimerization gaps exhibit the same
general behavior for the Hubbard and the $t-J$ model. The response
both as a function of the dimerization amplitude and the correlation
strength are very similar and quite close in absolute values even if
the gaps are slightly larger in the $t-J$ model than in the Hubbard
one.  For the large $U$ asymptote in the Hubbard model, it is
noticeable that the DMRG procedure fails to gives good estimations of
the dimerization gaps. This failure has already been
observed~\cite{mariegap} and may find its origin in a lack of
numerical accuracy when the absolute values involved in the gap
calculations become too large.  For large-$U$ and large-$\delta$, Penc
and co-workers~\cite{2delta} have shown that the charge gaps are
expected to coincide with a $2*\delta$ asymptote. It can be seen from
figure~\ref{fig:gap} that this large-$U$ limit is rather well
verified even for intermediate $\delta$.  It is noticeable that in the
non-interacting limit the $t-J$ effective model exhibits the proper
exponential behavior as a function of the correlation strength and the
dimerization amplitude.

\subsection{Bond orders}

We computed the bond orders for the Hubbard model and $t-J$ model.
However one can expect that the elimination of the explicit reference to the DOC in the
wave-function will strongly affect the values of the bond order.
Indeed, over a $<i,j>$ bond, the different configurations that
contribute to the bond order are 
\begin{itemize}
\item the local one-electron doublets which are explicitly treated in
the $t-J$ model, and  contribute for $1/2$ for the bonding one 
and $-1/2$ for the anti-bonding one,
\item the local triplet which is explicitly treated in the $t-J$
model and has a zero contribution to the bond-order, 
\item the essentially neutral, ground state singlet which is
{\em effectively} treated in the $t-J$ model and has a zero contribution to
the bond-order when the DOC states are ignored, but has a real
contribution of ${4|t| \over \sqrt{U^2+16t^2}} = {2|tJ| \over t^2 +
J^2}$  when the DOC are taken into account,
\item the essentially ionic singlets that are ignored in the $t-J$
model since their contribution is very small in the wave function and
has a real contribution to the bond order of ${-4|t| \over
\sqrt{U^2+16t^2}} = {-2|tJ| \over t^2 + J^2}$ for the symmetric one and of $0$ for the antisymmetric one,
\item the three electrons doublets that are ignored in the $t-J$ model
because of their very low occurrence in the system wave function and
should have a contribution to the bond order of $1/2$ for the bonding
one and of $-1/2$ for the anti-bonding one.
\end{itemize}
One sees immediately that in order to have a reasonable evaluation of
the bond order in the $t-J$ model one should at least add an effective
term setting up to the correct value the contributions of the  quite 
probable local neutral singlets. We will therefore define the effective $t-J$
bond order as the true $t-J$ bond order plus the exact contribution of the
essentially neutral singlets, that is
\begin{eqnarray*}
  \label{bo} \hat p &=& {1 \over 2} \sum_{<i,j>} \sum_{\sigma} \left(
  a^{\dagger}_{i,\sigma} a_{j,\sigma} + a^{\dagger}_{j,\sigma}
  a_{i,\sigma} \right) \nonumber \\ && + {2|tJ| \over t^2+J^2}
  \sum_{<i,j>} S\!g^\dagger(i,j)S\!g(ij)
\end{eqnarray*}

\begin{figure}[h]
\begin{center} \vspace{2eM} 
\resizebox{5cm}{5cm}{\includegraphics{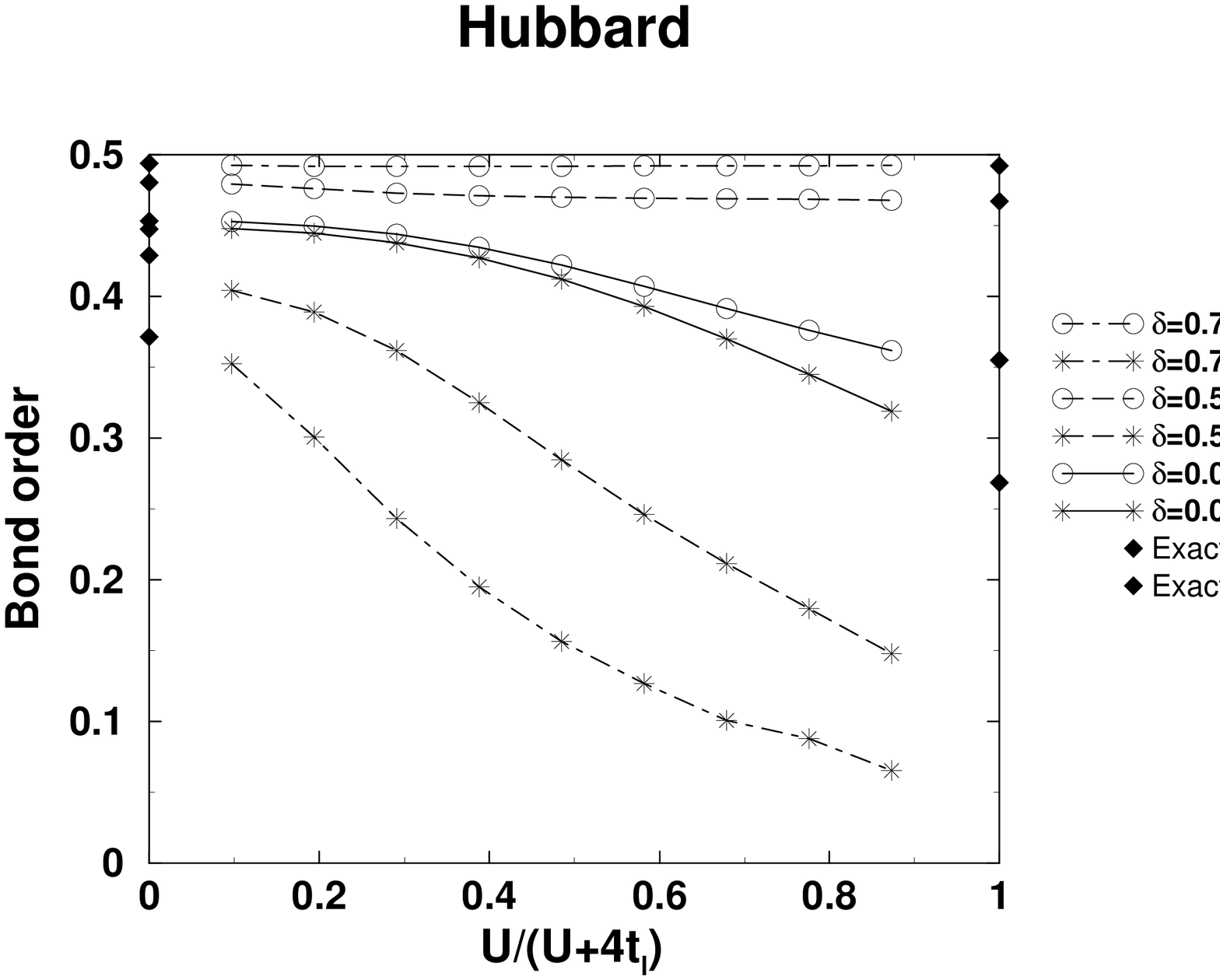}} \\ \vspace{2eM} 
\resizebox{5cm}{5cm}{\includegraphics{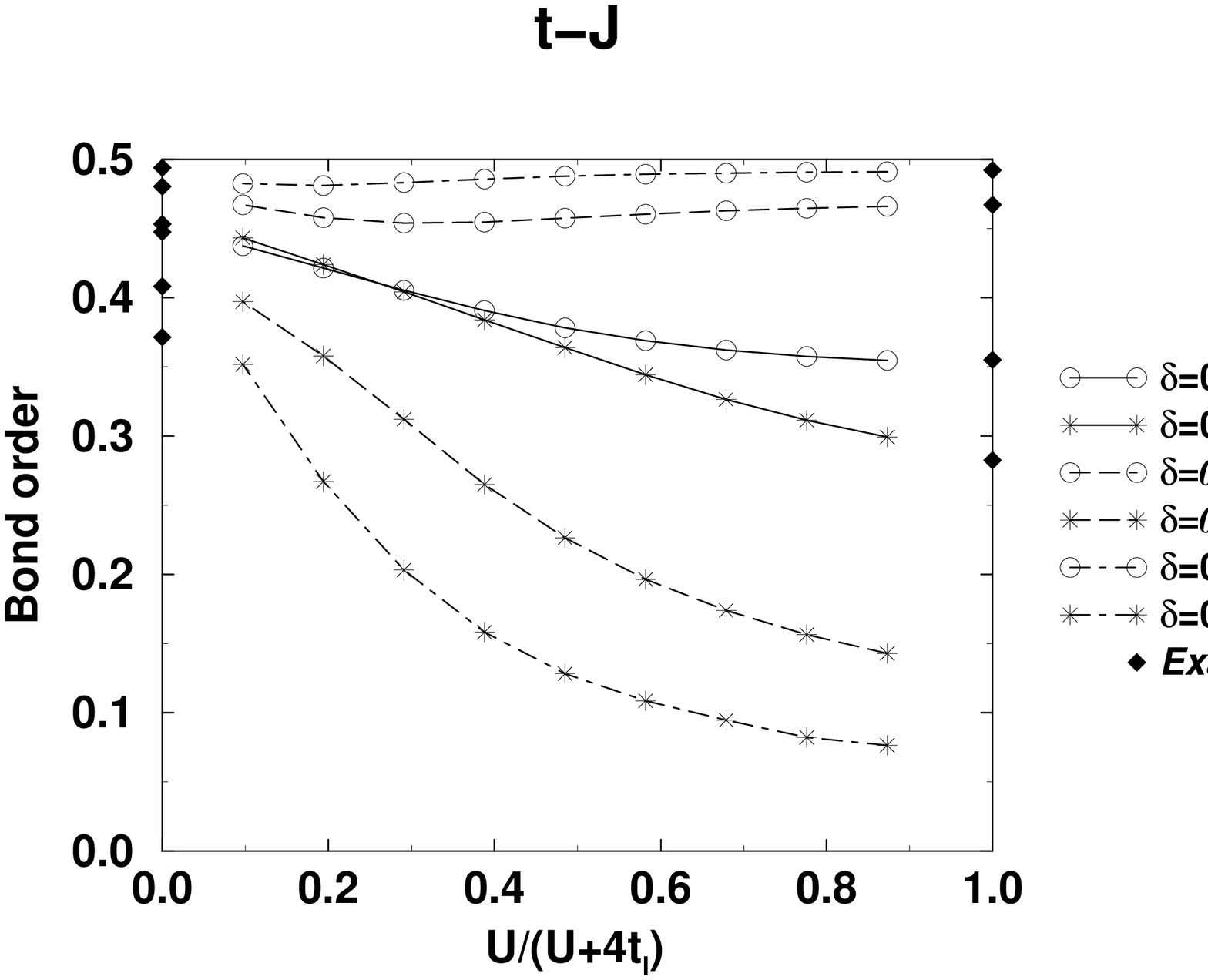}}
\caption{Bond orders $p$ calculated within the Hubbard model and
renormalized bond order calculated within the $t-J$ model. Circles
correspond to the short bonds (SB) and stars to the long bonds (LB).}
\label{fig:bo}
\end{center}
\end{figure}

Figure~\ref{fig:bo} shows the bond-order for the Hubbard model and the
effective bond-order for the $t-J$ model, for the whole range of
correlation strength and dimerization values.

The effective bond order of the $t-J$ model
behaves reasonably well both as a function of the correlation strength and
the dimerization. Both the strongly correlated limit and the $U=0$
limit of the Hubbard model are very well reproduced by the $t-J$
model. As for the energies, the major discrepancies between the two
models are in the moderately delocalized regime, typically between
$U/t=2$ and $U/t=4$, and for low dimerizations where the effective
$t-J$ bond order is slightly too rapidly decreasing.

One can conclude from the previous results that the $t-J$ model
reproduces quite well the low energy physics of the Hubbard model in
the low filling regime, over the {\em whole range} of the correlation
strength, provided that the exchange integral as well as the other
observables are defined so that to effectively take into account the
main effects of the DOC states.

\section{The three bodies terms}

From the perturbative expansion of the Hubbard model in the strong
correlation limit, we know that the exact effective model excluding
all DOC states should involve three, four, \dots, $n$ bodies
terms. After the dominant first neighbor effective exchange of the
$t-J$ model, second neighbor hopping and three bodies terms appear at
the second order of perturbation. These secondary terms which are
negligible near the complete or the very low fillings have a maximal
probability of occurrence at $1/3$ filling (at $U=0$). However their
relative importance compared to the local singlet or triplet terms
raises as $(1-\eta)^2$ as a function of the filling $\eta$.  In the
very low filling regime, these terms are therefore negligible, however
in the same way as the local singlets and triplets. In the low, but
not too low filling regime, one can expect that these terms may
account for some secondary corrections. Indeed, in the $1/4$-filling
regime used to exemplify this work, they have (for $U=0$) a total
probability of occurrence of $0.0198$.  One can therefore expect that
these terms may account for a large amount of the small quantitative
discrepancies observed between the Hubbard and effective $t-J$
models. Therefore, if one would like a more quantitative 
representation of the Hubbard model on the states excluding all double
occupancies on a site, one should  add to the $t-J$
Hamiltonian a second neighbor hopping and a three body term.
\begin{eqnarray*} \label{eq:3t}
H_{3b} &=& \; t \sum_{<i,j>} \sum_{\sigma} P\left(
    a^{\dagger}_{i,\sigma} a_{j,\sigma} + a^{\dagger}_{j,\sigma}
    a_{i,\sigma} \right)P \\&& - J/2 \sum_{<i,j>}
  P(a^{\dagger}_{i,\uparrow}a^{\dagger}_{j,\downarrow} -
    a^{\dagger}_{i,\downarrow}a^{\dagger}_{j,\uparrow})
  (a^{}_{i,\uparrow}a_{j,\downarrow} -
    a_{i,\downarrow}a_{j,\uparrow})P \nonumber \\ && 
 + t_2 \sum_{<i,j,k>} \sum_\sigma{ P\left(
a^\dagger_{i,\sigma} a_{k,\sigma} + a^\dagger_{k,\sigma} a_{i,\sigma}
\right)n_{j,\bar{\sigma}}(1-n_{j,\sigma})P} \\ && 
+ K \sum_{<i,j,k>}P{\left( a^\dagger_{i,\sigma} a_{k,\bar{\sigma}} +
a^\dagger_{k,\sigma} a_{i,\bar{\sigma}}\right)
a^\dagger_{j,\bar{\sigma}}a_{j,\sigma}(1-n_{j,\bar{\sigma}}) }P \nonumber
\end{eqnarray*}
where the sum over $<ijk>$ runs over all three nearest neighbor sites.
As for the effective exchange $J(t,U)$ of the $t-J$ Hamiltonian, the 
$J$, $t_2$ and $K$ parameters of $H_{3b}$ can be variationally derived
as a function of Hubbard $t$ and $U$, from the low energy spectroscopy
(between the six essentially neutral states) of the three
sites, two electrons fragment. It comes
\begin{eqnarray*}
J &=& {U\over 2} - {\sqrt{U^2+16t^2} + \sqrt{U^2+8t^2} \over 4} \\
t_2 = -K &=& {\sqrt{U^2+16t^2} - \sqrt{U^2+8t^2} \over 8}
\end{eqnarray*}

It should be noted that we are no more in a $t-J$ model and that the
free electrons system now maps onto a three body model with parameters
$J=(1+\sqrt{2})\,|t|$, $t_2=(1-\sqrt{2})/2\, |t|$ and
$K=-(1-\sqrt{2})/2\,|t|$.

\section{Conclusion}

The $t-J$ model is a well known effective Hamiltonian of the Hubbard
model in the strongly correlated limit. We have shown in this paper
that the $t-J$ model can be seen as such, as long as the double
occupancies of a site have a very low probability of occurrence in the
ground and low excited states of the system. This is in particular the
case for low filled bands (or nearly filled band using electron-hole
symmetry), whatever the value of the correlation strength. We have
proposed a variational evaluation of the effective exchange $J(t,U)$
so that it insures the proper excitation energy between the local
singlets and triplets. Using this effective $t-J(t,U)$ model, we have
shown on an infinite quarter filled chain, that the total energy per
site, the charge gaps as well as the bond-orders behave similarly than
in the Hubbard model, over the whole range of correlation strength and
dimerization. The agreement is not only qualitative, but also
quantitative, the small discrepancies seen in the intermediate low
(but not too low) correlation regime being analyzed as due to the
neglected three bodies terms. It is interesting to point out that the
super-symmetric point of the $t-J$ model corresponds to the effective
representation of the uncorrelated limit, $U=0$, yielding as a natural
consequence to the effective Hamiltonian theory, its surprising Fermi
liquid behavior.

\end{document}